\documentstyle[prl,aps,twocolumn]{revtex}
\begin{document}
\draft

\title{Peak Effect in Twinned Superconductors}

\author{A.I. Larkin$^{1,2}$, M.C. Marchetti$^{3}$, and V.M. Vinokur$^{1}$}
\address{$^1$ Argonne National Laboratory, Materials Science Division,
Argonne,
IL 60439}
\address{$^2$ L.D. Landau Institute for Theoretical Physics, 117940
Moscow, Russia}
\address{$^3$ Physics Department, Syracuse University, Syracuse, NY 13244}

\date{\today}

\maketitle

\widetext
\begin{abstract}
A  sharp maximum in the critical current $J_c$ as a function of temperature
just below the melting point of the Abrikosov flux lattice
has recently been observed in both low and high temperature
superconductors. This peak effect is strongest in twinned
crystals for fields aligned with the twin planes.
We propose that this peak
signals the breakdown of the collective pinning regime and the crossover
to strong pinning of single vortices on the twin boundaries.
This crossover is very sharp and can account for
the steep drop of the differential resistivity observed in experiments.

\end{abstract}

\narrowtext

The discovery of the high-temperature copper-oxide superconductors has renewed
the
experimental and theoretical
interest in the properties of the mixed state of type-II
superconductors in a magnetic field \cite{tora}.
Various experimental techniques, including standard current
versus voltage curves, are used to measure the critical current density
$J_c$ needed to depin the flux-line array and to investigate its temperature
and field dependence.
Naively $J_c$ is expected
to decrease monotonically as the temperature or the applied field
are raised towards the mean field $H_{c2}(T)$ line.
It has, however, been known for some time that
an abrupt increase in $J_c$ as a function of field or temperature
can occur in conventional low temperature superconductors near $H_{c2}$
\cite{kerchner}.
A qualitative explanation of this phenomenon, referred to as ``peak effect'',
was proposed a long time ago by Pippard \cite{pippard}, who argued that the
increase
in $J_c$ is associated with the softening of the
shear modulus $c_{66}$.
A more quantitative explanation of the peak effect
as arising from the softening of all the elastic
moduli of the flux lattice  near $H_{c2}$ was presented
by Larkin and Ovchinnikov
\cite{lo}.

Recently a sharp maximum in $J_c$ as a function of  temperature has been
observed in both untwinned \cite{budnick,danna,kwokpriv,yeh} and twinned
\cite{kwokpeak}
YBCO crystals, as well as in some low temperature superconductors
\cite{shobo,ak}. The new feature is that in this case the peak occurs
away from $H_{c2}$, just below the temperature
$T_m$ where the flux lattice
melts
into a flux-line liquid. In view of the old suggestion by Pippard
\cite{pippard} and the
work by Larkin and Ovchinnikov \cite{lo}, it is natural to associate it with
the softening
of the shear modulus $c_{66}$ that occurs at the melting point.
In twinned $YBCO$ crystals the peak depends strongly on the orientation of
the applied field relative to the twin planes: it is largest for flux motion
along the twin planes and external fields
aligned with the $c$ axis and it weakens as the field
is tilted out of the plane of the twins \cite{kwokpeak}.
In untwinned
YBCO single crystals the peak is much smaller: it shifts
towards $T_c$ and
becomes less pronounced as the sample
\newline\vskip 1.27truein\noindent
purity is increased \cite{danna}.
These observations suggest that strong anisotropic pinning centers such as
twin boundaries
enhance the peak effect.

In this paper we first examine in detail the temperature
dependence of the
critical current from collective pinning by
point defects \cite{fglv,fv,tora},
and show that  $J_c$  can exhibit a sharp rise near $T_m$
for a narrow range of magnetic fields
due to the abrupt decrease of the shear modulus.
This may provide a mechanism for the small peak effect observed in
untwinned single crystals.
New results on anisotropic
collective pinning in samples with
a family of parallel twin planes are also presented and show that
the same mechanism is in principle operative in twinned samples,
when vortices are pinned collectively by an array of twin planes.
On the other hand,
collective pinning is very weak in this case and
cannot account for the large increase in $J_c$ observed in these samples.
In the second part of the paper we show that in twinned samples
a large peak in $J_c$
near melting can arise from
strong pinning of individual vortices on the twins.
This mechanism is enhanced at $T_m$ by
the vanishing of $c_{66}$ and can account for the sharp drop in the
resistivity observed in experiments.

The critical current density of an elastic medium pinned by weak
disorder can be calculated using the collective pinning theory \cite{lo}.
Weak disorder
destroys the translational order of the flux-line lattice and
results in the coherent pinning of vortex bundles of extent
$L_c$ and $R_c$ in the directions
parallel and perpendicular to the applied field,
${\bf H}$.
The pinning lengths $R_c$ and $L_c$, defined as the distances at
which the lattice distortion
due to disorder is of the order of the range $\xi$ of the pinning potential,
are determined in terms of the elastic constants of the lattice
by balancing the elastic deformation energy against the
pinning energy.
The critical current $J_c$ is then defined as the current where
the Lorentz force balances the pinning force, or
$BJ_c/c\approx\sqrt{W/V_c}$, where $W=n_p<f^2>$ is the mean
square pinning force,
with $n_p$ the volume density of
pins and $f$ the elementary pinning force, and $V_c=R_c^2L_c$.

We consider a three dimensional flux-line array in
a sample with an external magnetic field aligned with the
$c$ axis, which is chosen as the $z$ direction.
Disorder
is described as a quenched random potential per unit length
$V({\bf r} )$ with zero
mean and Gaussian correlations,
$\overline{V({\bf r} )V({\bf r} ')}=\Gamma({\bf r} ,{\bf r} ')$.
The overbar denotes the disorder average and
the correlator $\Gamma({\bf r} ,{\bf r} ')$ is determined by the strength and
geometry of the disorder.
The static elastic deformation of the lattice due to disorder can
be evaluated by a perturbation theory in the pinning potential
\cite{fglv,tora}.
To lowest order in perturbation theory  the components of the mean square
displacement
$U_{ij}({\bf r} )=<\overline{\Delta u_i({\bf r} )\Delta u_j({\bf r} )}>$
induced by the random potential
(here $\Delta u_i({\bf r} )=u_i({\bf r} )-u_i(0)$ and the brackets denote
a thermal average) are given by \cite{fglv,tora},
\begin{eqnarray}
\label{eq:meansq}
U_{ij}({\bf r} )=& &\int{d{\bf q} \over(2\pi)^3}
\int{d{\bf q} '\over(2\pi)^3}(1-e^{i{\bf q} \cdot{\bf r} })(1-e^{i{\bf q}
'\cdot{\bf r} '})\nonumber\\
& &G_{ik}({\bf q} ,\omega=0)G_{jl}({\bf q} ',\omega=0)
\tilde{W}_{kl}({\bf q} +{\bf q} '),
\end{eqnarray}
with $G_{ij}({\bf r} ,t)$ the elastic Green function of
the lattice,
${\bf f}_p({\bf r} )=-n({\bf r} ,t)\vec{\nabla}V({\bf r} )$
the pinning force per unit volume and
$\tilde{W}_{kl}({\bf q} ,{\bf q} ')=\overline{f_{p,i}({\bf q} )f_{p,j}({\bf q}
')}$
the pinning force correlator.
Here $\hat{n}({\bf r} ,t)=\sum_n\delta^{(2)}({{\bf r}_{\perp}}-{\bf r}
_n(z,t))$,
with ${\bf r} =({{\bf r}_{\perp}},z)$,
is the coarse-grained microscopic vortex density field
(the flux lines are parametrized by their trajectories
$\{{\bf r}_n(z,t)\}$).
The main contribution to Eq. (\ref{eq:meansq}) for the case of interest below
comes from the
transverse part of the elastic Green's function, given by
\begin{equation}
\label{eq:green}
G^T_{ij}({\bf q} ,\omega)={{\cal P}^T_{ij}\over -i\omega\zeta
+c_{66}{q_\perp} ^2+c_{44}({q_\perp} ,q_z)q_z^2},
\end{equation}
where ${\cal P}^T_{ij}=\delta_{ij}-\hat{q}_{\perp i}
\hat{q}_{\perp j}$, with $\hat{q}_{\perp i}=q_{\perp i}/q_{\perp}$,
and
$c_{66}$ and $c_{44}({q_\perp} ,q_z)$ are the shear and tilt moduli of the
vortex lattice, respectively.
Thermal fluctuations can be incorporated in the perturbation theory
by separating out in the vortex positions ${\bf r}_n(z,t)$
the deviation from equilibrium due to pinning from that due to thermal
effects, as described in \cite{fv}. The main effect of thermal
fluctuation
is the replacement
of the upper cutoff $q_0\approx\xi^{-1}$ in the wavevector
integral by a thermal cutoff $q_T\approx(\xi^2+<u^2>_{th})^{-1/2}
=q_0(1+T/T_{dp})^{-1/2}$. The depinning temperature  $T_{dp}$ is
defined by $<u^2(T_{dp})>_{th}\approx\xi^2$,
where $<u^2(T)>_{th}$ is the mean square thermal excursion
of the vortices about their equilibrium positions.
When the vortex array is described as an elastic continuum,
$T_{dp}=2a_0\xi^2\sqrt{c_{66}c_{44}}
=2a_0^2\xi^2\sqrt{c_{66}\hat{c}_{44}}/\lambda$,
with $a_0=\sqrt{\phi_0/B}$ the mean intervortex separation,
$\lambda$ the penetration length in the $ab$ plane and
$\hat{c}_{44}=c_{44}({q_\perp} =0,q_z=0)$ \cite{notea}.

When the flux array is pinned by isotropic point disorder, the random
potential
is short-ranged in all directions, with $\Gamma({\bf r} ,{\bf r} ')
=\gamma\delta^{(3)}({\bf r} -{\bf r} ')$ and
$\gamma\approx(U_0\xi^3)^2n_p\big[1+{\cal O}(n_p\xi^3)\big]$, with
$U_0$ the depth of an individual
pinning potential. The pinning force correlator
is isotropic,
$W_{ij}({\bf q} )=W\delta_{ij}(2\pi)^3\delta^{(3)}({\bf q} )$, with
$W=\gamma/\xi^4a_0^2$.
The pinning lengths and the critical current
for this disorder geometry have
been calculated elsewhere \cite{fglv}, but it is instructive
to display the dependence of $J_c$ on the elastic constants.
For low defect densities ($R_c>\lambda$ - this is the so-called large
bundles regime), the
dispersion of the tilt modulus can be neglected and
$J_c\approx j_0({j_{sv}\over j_0}{H_c^3\over 4\pi})^3{3\sqrt{3}B\over 16
H_{c2}}{1\over c_{66}^2\hat{c}_{44}}
\Big(1+T/T_{dp}\Big)^{-11/2}$.
Here $j_0=cH_c/3\sqrt{6}\pi\lambda$ is the depairing current
and $j_{sv}=j_0\big(Wa_0^2\xi/\epsilon_0^2\big)^{2/3}$ is the single
vortex critical current density.
The critical current contains both an explicit temperature dependence
from the thermal smoothing of the pinning potential
and an implicit $T$ dependence through the superconductor's
parameters that determine the elastic constants.
To extract the strong temperature dependence of $c_{66}$ near melting,
we write $c_{66}=c_{66}^0r(T/T_m)$, where $c_{66}^0$
depends weakly on $T$ and the function  $r$
drops sharply from unity to zero at the clean flux lattice
melting temperature $T_m$. This is defined by
$<u^2(T_m)>_{th}\approx c_L^2a_0^2$, with $c_L\approx 0.1-0.3$ the Lindemann
parameter.
The critical current is then $J_c\sim {1\over c_{66}^2r^2(T)}
\Big(1+T/T_{dp}^0r^{1/2}(T)\Big)^{-11/2}$, where
$T_{dp}^0=2a_0^2\xi^2\sqrt{c_{66}^0\hat{c}_{44}}/\lambda$.
The temperature dependence of $J_c$ near $T_m$ is controlled by the
parameter $\alpha=T_m/T_{dp}^0=(c_La_0/\xi)^2$.
For $H=6T$ and temperatures near
$T_c$, we use $\xi\approx 100\AA$ and $c_L\approx 0.2$
to obtain $T_m/T_{dp}^0\approx 0.1$.
At low temperatures ($T<<T_{dp},T_m$)
$J_c$ decreases very slowly with $T$.
At higher temperatures, but still well below $T_m$,
the elastic constants are only very weakly temperature
dependent and the temperature dependence of $J_c$ is controlled by
thermal fluctuations, yielding a decrease of $J_c$ as $T$ grows.
As $T_m$ is approached from below, $c_{66}$ softens
and the flux lattice can better adjust to the pinning centers, raising
$J_c$. Finally, near $T_m$ the function $r(T)$ and therefore $T_{dp}$
drop sharply in a narrow temperature range
giving rise to an increase of the pinning barriers and
an associated drop in $J_c$.
For larger defect densities the dispersion of
$c_{44}$ is important (this is referred to as the small-bundle regime).
In this case $J_c\approx j_0
{6\pi\sqrt{3}\over H_c^2} c_{66}
\Big(1+T/T_{dp}\Big)^{1/2}
\exp\Big[- ({j_0\over j_{sv}}{4\pi\over H_c^2})^{3/2}
{16\sqrt{2\pi}\over B}\hat{c}_{44}^{1/2}
c_{66}^{3/2}\big(1+T/T_{dp}\big)^3\Big]$.
Again,
the
critical current can be written as
$J_c\sim \exp\{-ar^{3/2}[1+T/(T_{dp}^0r^{1/2})]^3\}$,
where $a$ is practically independent of temperature near $T_m$.
We find $J_c\sim e^{-a}$ below $T_m$ and
$J_c\sim e^{-\alpha^3 a}$ as $T\rightarrow T_m^-$,
yielding a sharp rise of $J_c$ in a very narrow temperature range.
This mechanism may be responsible for the small peak
effect observed in untwinned single crystals.  We remark, however,
that the melting
transition is broadened by the presence of defects and the drop in critical
current is more likely associated with the onset of plastic motion of
vortices.

We now consider collective pinning in a sample with a single family of
twin boundaries of mean separation $d$ spanning the $zy$ plane
for fields aligned with the twin planes. This is the experimental
geometry where the peak in $J_c$ is strongest \cite{kwokpeak}.
We define two different pinning
lengths, $R_{c\parallel}$ and $R_{c\perp}$, corresponding to the size
of the vortex bundle in the directions parallel and transverse to
the twin planes, respectively. If $R_{c\perp}>>d$, pinning
occurs via the collective action of many twin planes. Each twin
is described as a sheet with a large concentration of point defects.
The correlator of the random potential is given by
$\Gamma({\bf r} ,{\bf r} ')=\gamma_1 g(|x-x'|)\delta(y-y')\delta(z-z')$, where
$\gamma_1\approx (U_0\xi^3)^2n_p^{(2)}\big[1+{\cal O}(\xi/d)\big]$
is proportional to the {\it areal} density $n_p^{(2)}$ of pins
on each twin plane,
and $g(x)$ describes correlations
in the distribution of twin planes.
On distances large compared to the twin spacing $d$ the twins are
essentially uncorrelated [$g(x)\approx (1/d)\delta (x)$] and the
pinning force correlator is
$\tilde{W}_{ij}({\bf q} +{\bf q} ')=W_T
\delta_{ij}(2\pi)^3\delta^{(3)}({\bf q} +{\bf q} ')$,
with $W_T=\gamma_1/\xi^4da_0^2$.
Collective pinning in this regime is very similar to collective
pinning by point defects in bulk.
The dependence of $J_c$ on the elastic constants and temperature
is identical to that obtained for isotropic point disorder, with
the replacement $W\rightarrow W_T$.
A peak effect in densely twinned samples may then in principle
arise from the same
mechanism discussed above for untwinned crystals.
On the other hand, the mean squared pinning force $W_T$ is still determined
by the effective {\it volume}
density of pins, which is now given by $n_p^{(2)}a_0/{\xi}d$.
As a result, the anisotropy due to the twin planes increases
the pinning volume of a factor $(d/d_p)^3$, with
$d_p\sim(n_p)^{-1/3}$,
correspondingly
decreasing
the critical current. For this reason collective pinning in twinned
samples is very weak, especially if $d>>a_0$ and cannot account for
the observed critical currrents.

The dominant pinning mechanism in twinned crystals,
particularly in sparsely twinned samples,
is the strong
pinning of individual vortex lines trapped on the twin boundaries.
As $T_m$ is approached fom below intervortex interactions weaken
and the vortices on the twins become more strongly pinned than those
in the channels between twins.
The main contribution to the critical current arises then
from pinning of single vortices on the twins, and particularly from those
vortex segments that are strongly pinned in rare regions with an excess
of impurities. As described below, it is the rise in the fraction of
such strongly
pinned vortex segments on the twins with $T$
that can be responsible for the peak
in $J_c$ in twinned samples.

To evaluate the critical current due to strong pinning in regions
with excess impurities, we consider
a representative vortex line trapped near a twin plane and
interacting with its neighbors at an average distance $a_0$
in the lattice. The remainder of the lattice, even though not directly pinned
by the twin, is held in place by interactions. The magnitude of the
elastic force
associated with displacing a length $L$
of the representative fluxon a transverse distance $u$ from its
equilibrium position is
\begin{equation}
\label{eq:svcross}
F_{el}(u,L)\sim \tilde{\epsilon}_1{u\over L}
+c_{66}u L.
\end{equation}
The first term is
the force associated with tilting the representative
vortex, with $\tilde{\epsilon}_1$ the tilt energy per unit length.
The second term arises from the interaction with the neighbors.  The
typical pinning force exerted on a vortex segment of length L is
$ (\overline{F_p^2(L)})^{1/2}\sim (W_1\xi^2L)^{1/2}$,
with $W_1=\gamma_1/\xi^5a_0^2$ the mean squared pinning force
per unit volume due to a single twin of thickness $\xi$.
The most effective pinning
arises
from rare regions with an anomalously large impurity
concentration that pin strongly the vortex segment.
The pinning forces $F_p$ in these regions exceed the typical pinning
force $ (\overline{F_p^2(L)})^{1/2}$ and give the dominant contribution
to $J_c$.
The problem of strong pinning of vortex lines
is analogue to that of incommensurate charge density
waves and can be rigorously discussed following Ref. \cite{vmf}.
Here we prefer, however, to follow the more phenomenological, but physically
intuitive discussion given by Coppersmith \cite{coppersmith}.  The condition
for the strong pinning is $F_p(\xi)>F_{el}(\xi)$ \cite{lo,labusch}.
The critical current
$J_c$ is proportional to the density $n$ of the strongly pinned vortex
segments where
this condition is satisfied.  To find it we note that the pinning force $ F_p$
scales as the impurity excess in the region and it can be shown to be
Gaussian-distributed with variance $(\overline{F_p^2})^{1/2}$
\cite{coppersmith}.
The density $n$ of vortex segments that are strongly pinned
in these excess-impurity regions is then,
\begin{equation}
\label{eq:strong}
n\sim\int_0^\infty dL \int_{F_{el}(L)}^\infty dF_p e^{-(F_p)^2/2
(\overline{F_p^2(L)})},
\end{equation}
where we have used $u\sim\xi$. Using Eq. \ref{eq:svcross},
one can show that the integral over $L$ is dominated
by the length scale $L^*\sim\sqrt{\tilde{\epsilon}_1/c_{66}}$
where the single-vortex tilting force balances the force of interaction
of the pinned vortex with the rest of the lattice.
The critical current $J_c$ is then given by,
\begin{equation}
\label{eq:strongp}
J_c\sim n\sim\exp[-(c_{66}/c^*_{66})^{3/2}],
\end{equation}
where $c^*_{66}=(2W_1/\sqrt{\tilde{\epsilon}_1})^{2/3}$.
As $T$ approaches $T_m$, the shear modulus softens and drops to zero.
Correspondingly, $J_c$ grows according to Eq. (\ref{eq:strongp}).
The result given in Eq. (\ref{eq:strongp}) applies for $c_{66}\geq c^*_{66}$.
The mechanism of strong pinning just described is
also operative in untwinned samples where vortices are pinned by
isotropic point disorder. We note that strong pinning of isolated
vortices yields the same functional dependence of $J_c$ on $c_{66}$
as collective pinning of
small bundles. In both cases $J_c\sim\exp(-ac_{66}^{3/2})$.
In untwinned crystals $a$ is essentially the same for these two
pinning mechanisms.
In twinned crystals
collective pinning of vortex bundles is very weak ($a\sim 1/W$)
and strong single
vortex pinning in regions with excess impurities ($a\sim 1/W_1$)
controls the critical
current even
when the condition for single-vortex pinning is satisfied only
locally.
In this case the main contribution to $J_c$ arises from
pinning energy
barriers which are large compared to the typical barrier
$E_p(L^*)\sim(W_1\xi^2/c^*_{66})\big(c^*_{66}/c_{66}\big)^{1/4}$,
but small compared to the scale of the elastic energy of interaction
of the vortex segment with the rest of the lattice
$c_{66}\xi L^*\sim \sqrt{\tilde{\epsilon}_1 c_{66}}\xi\sim
(W_1\xi^2/c^*_{66})\big(c_{66}/c^*_{66}\big)^{1/2}$.

Most experiments do not measure the critical
current as defined theoretically, but rather the nonlinear resistivity.
For comparison with experiments it is important to
discuss the small finite resistivity
due to the creep of strongly pinned vortex segments
at low measuring currents, $J<J_c$.
To find the energy barriers that determine the creep rate,
we consider the distribution of barriers separating different metastable
states in the regime of strong pinning. As discussed above for
pinning forces, it has been shown that the impurity energies
$\delta E$
of strongly pinned vortex segments of length $L$
are described by a
Gaussian distribution,
$\sim\exp[-(\delta E)^2/2W_1L\xi^4]$ \cite{coppersmith}. A gain
$\delta E$ in pinning energy is associated with a cost
$\delta{\cal F}_{el}=\tilde{\epsilon}_1\xi^2/L+c_{66}\xi^2L$
in elastic energy.
The resulting activation barrier is therefore
$E_p=\delta E-\delta{\cal F}_{el}$.
The distribution function of creep barriers can be evaluated following
\cite{coppersmith},
\begin{equation}
\label{eq:creepbar}
P(E_p)\sim\int_0^\infty dL e^{-(E_p+\delta{\cal F}_{el})^2/
2W_1L\xi^4}\sim e^{-{2E_pc_{66}\over W_1\xi^2}}.
\end{equation}
The typical creep barrier at $J \sim J_c$ is
$E_p\sim W_1\xi/c_{66}$,
At arbitrary currents the creep barrier becomes $E_p(J)\sim
W_1\xi/c_{66}f(J/J_c)$.  The function $f(J/J_c)$
decreases as $J/J_c$ grows, but its explicit form  cannot be
obtained by our dimensional analysis.
The resulting thermally
activated resistivity is
\begin{equation}
\label{eq:creep}
\rho\sim {1\over J_c}\exp\Big(-{W_1\xi^2f(J/J_c)\over Tc_{66}}\Big),
\end{equation}
with $J_c$ given by Eq. (\ref{eq:strongp}).  The temperature
dependence of the resistivity
at a fixed current is governed by the
temperature dependence of $J_c$.
When $c_{66}$ decreases near melting, the creep resistivity drops
exponentially. This result applies for
$J\leq J_c$ and temperatures near, but below melting, where $c_{66}$
is not too small, $c_{66}>\sqrt{W_1\xi}$.

In the regime just described dominated by strong pinning of individual
vortex segments on twin planes, interactions are still strong enough
to hold the lattice
together so that the remainder of the flux array can be described
as an elastic continuum.
On the other hand, in the presence of an applied current
the competition between
strong single-vortex pinning at the twin boundaries and the elastic
deformations of the portions of flux lattice between twin planes
eventually leads to the development of large strains in the regions next
to the twins. This results in the break-up of the lattice and
the onset of plastic flow.
The condition for the onset of plastic flow is obtained by considering
the elastic deformation of a flux lattice in the channel
$0\leq x\leq d$ between
two twin planes driven by
a current transverse to the twins,
yielding a Lorentz force $f_L=BJ/c$ per unit volume in the $+y$ direction.
The displacement $u_y(x)$ of the flux lattice is the solution
of $\zeta \partial_tu_y=c_{66}\partial^2_xu_y+f_L$, with boundary conditions
$u_y(0)=u_y(L)=0$. In an elastic continuum the mean velocity
$v_y=\partial_tu_y$ must be spatially homogeneous
and $u_y(x)=(f_L/2c_{66})x(d-x)$.
The lattice yields when the strains at the twin boundaries become
too large and the condition  $|\partial_x u_y(x)|<< 1$ is no longer
satisfied. An approximate condition for the onset of plastic flow
can then be written as $f_Ld\sim c_{66}$, with a corresponding
current scale, $J_p\sim c_{66}/Bd$.
The onset of plastic flow at $J\sim J_p$ corresponds to a sharp rise in
the differential resistivity and the critical current.

\vskip .2in
This work was supported by the National Science Foundation at Syracuse
through Grants No. DMR91-12330, DMR92-17284 and DMR94-19257,
at the ITP of the University of California in Santa Barbara through
Grant No. PHY89-04035, and through the U.S. Department of Energy,
BES-Material Sciences, under contract No. W-31-109-ENG-38.


\begin{thebibliography}{99}
\bibitem{notea}
Even though only the long wavelength tilt modulus enter in this expression,
we emphasize that
$T_{dp}$ is mainly determined by wavevectors
$q\simeq a_0^{-1}$.
\end{thebibliography}
\end{document}